\documentclass[preprints,article,accept,pdftex,moreauthors]{Definitions/mdpi} 
\firstpage{1} 
\pubvolume{1}
\issuenum{1}
\articlenumber{0}
\pubyear{2023}
\copyrightyear{2023}
\datereceived{ } 
\daterevised{ } 
\dateaccepted{ } 
\datepublished{ } 
\hreflink{https://doi.org/} 

\Title{Heavy flavor transport and observables in heavy-ion collisions within the MARTINI+MUSIC framework}

\Author{Manu Kurian $^{1, 2, *}$\orcidA{}, Mayank Singh $^{3, 4}$\orcidB{}, Sangyong Jeon $^{1}$\orcidC{} and Charles Gale $^{1}$\orcidD{}}

\AuthorNames{Manu Kurian, Mayank Singh, Sangyong Jeon and Charles Gale}

\AuthorCitation{Kurian, M.; Singh, M.; Jeon, S.; Gale, C.}

\address{%
$^{1}$ \quad Department of Physics, McGill University, 3600 University Street, Montreal, QC, H3A 2T8, Canada\\
$^{2}$ \quad RIKEN BNL Research Center, Brookhaven National Laboratory, Upton, New York 11973, USA\\
$^{3}$ \quad School of Physics and Astronomy, University of Minnesota, Minneapolis, MN 55455, USA\\
$^{4}$ \quad Department of Physics \& Astronomy, Vanderbilt University, Nashville, TN 37235, USA
}

\corres{Correspondence: mkurian@bnl.gov}

\abstract{We study the transport dynamics of charm quarks within an expanding quark-gluon plasma for Pb+Pb collisions at 2.76 TeV. The analysis incorporates the hydrodynamical approach-MUSIC with fluctuating IP-Glasma initial state and Bayesian-quantified viscous coefficients. We study the interaction strength of charm quarks in the medium, including elastic collisional processes with medium constituents, gluon emission processes, and the impact of non-perturbative interactions on heavy quark transport. Further, we analyze the dynamics of heavy flavors using a hybrid framework that incorporates the MARTINI event generator, with PYTHIA8.1 for the initial production of heavy quarks, and Langevin dynamics to describe the evolution of heavy quarks.}


\begin{document}

\section{Introduction}
Owing to the complexity of the nuclear reaction, the formation and evolution of the strongly interacting matter – the quark-gluon plasma (QGP) created at the heavy-ion collision experiments are modeled in stages. These multistage models encompass the initial stage dynamics, the fluid dynamical evolution of the QGP, and the subsequent phase involving hadronic gas. Heavy quarks, predominantly created in the early collision stages, act as witnesses to QGP evolution because of their relatively longer thermalization time compared to the QGP's lifespan. The Brownian dynamics of heavy flavor particles within the QGP medium can be thoroughly explored within the Langevin framework or Boltzmann equations. The interaction of heavy quarks within the QGP medium can be characterized by their drag and diffusion coefficients. These coefficients offer a quantitative description of how heavy quarks experience interactions and movement within the QGP environment. Analytical estimation of heavy quark transport coefficients has traditionally relied on the assumption of a medium in local thermal equilibrium. However, it is well-established that substantial deviations from thermal equilibrium persist throughout all phases of heavy-ion collisions. In the current analysis, we study the charm quark transport in an expanding medium. The space-time evolution of the QGP is examined using a realistic hydrodynamical approach, specifically MUSIC~\cite{Schenke:2010nt}, with a fluctuating IP-Glasma initial state~\cite{Schenke:2012wb}. Further, the dynamics of the charm quark are modeled with MARTINI~\cite{Schenke:2009gb} to analyze heavy flavor observables in heavy-ion collisions.
\section{Formalism}
The dynamics of heavy quark in the QGP medium depend upon its interactions with the constituents in the medium. The Langevin equation provides a standard approach for studying the motion of heavy quarks in a medium as,
\begin{align}
 dp_i=-A_i\, dt+C_{ij}\rho_j\sqrt{dt},
\end{align}
where $dp_i$ is the momentum shift over a time interval $dt$. The drag force $A_i$ can be defined as,
\begin{align}
&A_i=p_iA(p^2, T),
&&A=~\langle\langle1\rangle\rangle-{\langle\langle{\bf p}.{\bf p}^{'}\rangle\rangle}/{p^2},
\end{align}
with $A$ as the drag coefficient. The drag force characterizes the average momentum transfer experienced by heavy quarks due to interactions, while the matrix $C_{ij}$ quantifies the stochastic force using the Gaussian-normal distributed random variable $\rho_j$ \cite{Das:2013kea}. The dependence of heavy quark transport coefficients on  the temperature of the medium can be explored within relativistic transport theory. In the framework of the Fokker-Planck equation, the thermal average of a function $F(p')$ for the elastic collisional process, $HQ(P)+l(Q)\rightarrow HQ(P')+l(Q')$, where $l$ represents light quarks or gluons, can be written as,
\begin{align}\label{Ap1c}
\langle\langle F(|{\bf p}'|)&\rangle\rangle=\dfrac{1}{\gamma_{HQ}}\dfrac{1}{2E_p}\int{\dfrac{d^3{\bf q}}{(2\pi)^3 2E_q}}\int \frac{d^3 {\bf p'}}{(2 \pi)^3 2E_{p'}}\int \frac{d^3 {\bf q'}}{(2 \pi)^3 2E_{q'}}(2\pi)^4\delta^4(P+Q-P'-Q')\nonumber\\
&\times\sum|{\mathcal{M}}_{2\rightarrow 2}|^2 f_{g/q}({E_q})\Big(1 \pm f_{g/q}(E_{q'})\Big)F(|{\bf p}'|),
\end{align}
where $\gamma_{HQ}$ represents the degeneracy factor of the charm quark, $|{\mathcal{M}}_{2\rightarrow 2}|$ denotes the interaction strength of the charm quark-medium particles elastic scattering process, and $f_{g/q}$ is the distribution function of thermal particles in the evolving medium.
\section{Results}
\subsection{Impact of viscosity on charm quark energy loss}
As the evolution of the QGP medium is modeled within the viscous hydrodynamics, the medium is not exactly in thermal equilibrium. The near-equilibrium distribution can be defined as~\cite{Paquet:2015lta},
\begin{equation}\label{1.8}
f_{g/q}(Q, X)= f_{g/q}^0(Q)+\delta f_{g/q}^{{\text{shear}}}(Q, X)+\delta f_{g/q}^{{\text{bulk}}}(Q, X),
\end{equation}
where $f_{g/q}^0(Q)$ is the equilibrium momentum distribution, $\delta f_{g/q}^{{\text{shear}}}$ and $\delta f_{g/q}^{{\text{bulk}}}$ denote the shear and bulk viscous corrections to the distribution function, respectively.  We refer to Ref.~\cite{Kurian:2020orp} for detailed discussions on the shear and bulk viscous corrections to heavy quark transport coefficients for the elastic collisional process. 
\begin{figure*}
    \centering
    \includegraphics[scale=0.525]{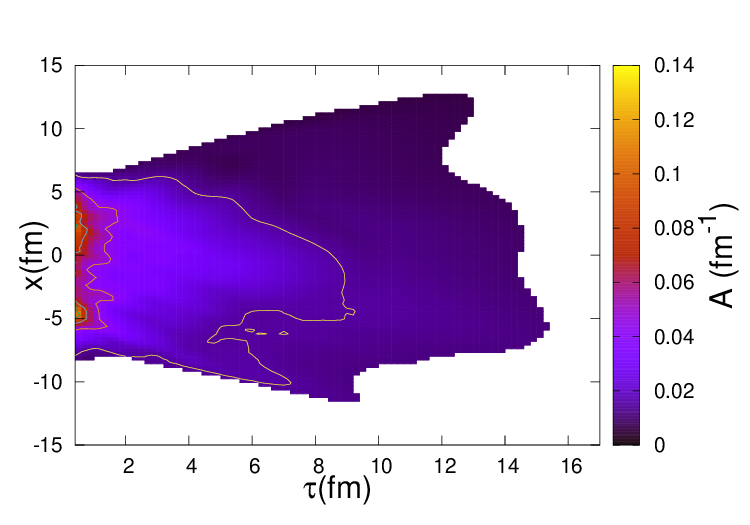}
      \includegraphics[scale=0.5]{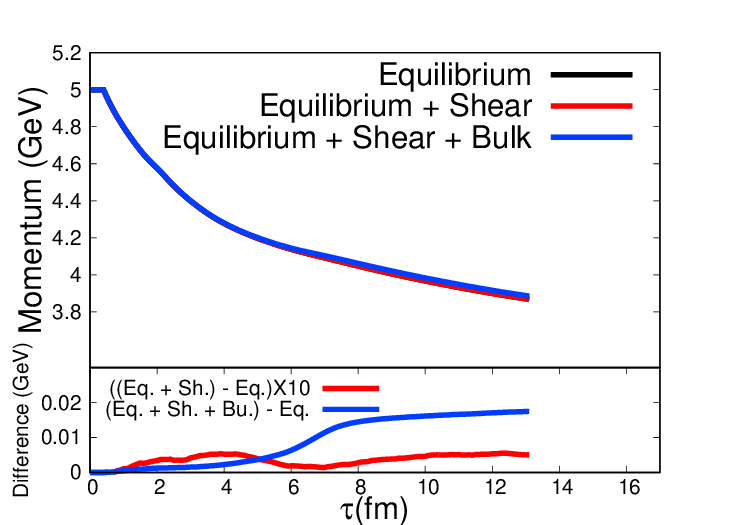}
    \caption{(Left panel) Charm quark drag coefficient at $p = 5$ GeV at different space-time points. (Right panel) Impact of viscous effects on charm quark momentum evolution in the QGP medium. }
    \label{fig1}
\end{figure*}
In Fig.~\ref{fig1} (left panel), the charm quark drag coefficient is plotted at different spacetime points for Pb+Pb collision at $\sqrt{s_{NN}}= 2.76$ TeV. We utilize the IP-Glasma model~\cite{Schenke:2012wb} to initialize the QGP and the viscous hydrodynamical approach MUSIC~\cite{Schenke:2010nt} for its evolution. It is observed that the drag coefficient decreases as the QGP expands in space-time, indicating that the medium offers less resistance to the motion of charm quarks at lower temperature regimes. Consequently, charm quarks experience more random forces in the early stage of the QGP evolution compared to its equilibrated stage, leading to a more random motion within the medium. The drag force, which resists the motion of heavy quarks, leads to its energy loss in the medium. This energy loss of the charm quark is illustrated by analyzing its momentum evolution in the expanding  QGP medium, as depicted in Fig.~\ref{fig1} (right panel). Notably, viscous corrections have minimal effects on the momentum evolution of the charm quark in the medium.
\begin{figure*}
    \centering
    \includegraphics[scale=0.5]{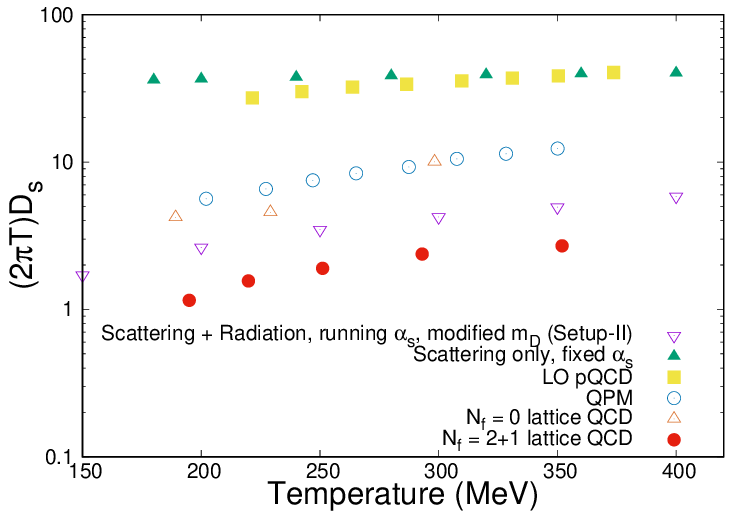}
      \includegraphics[scale=0.155]{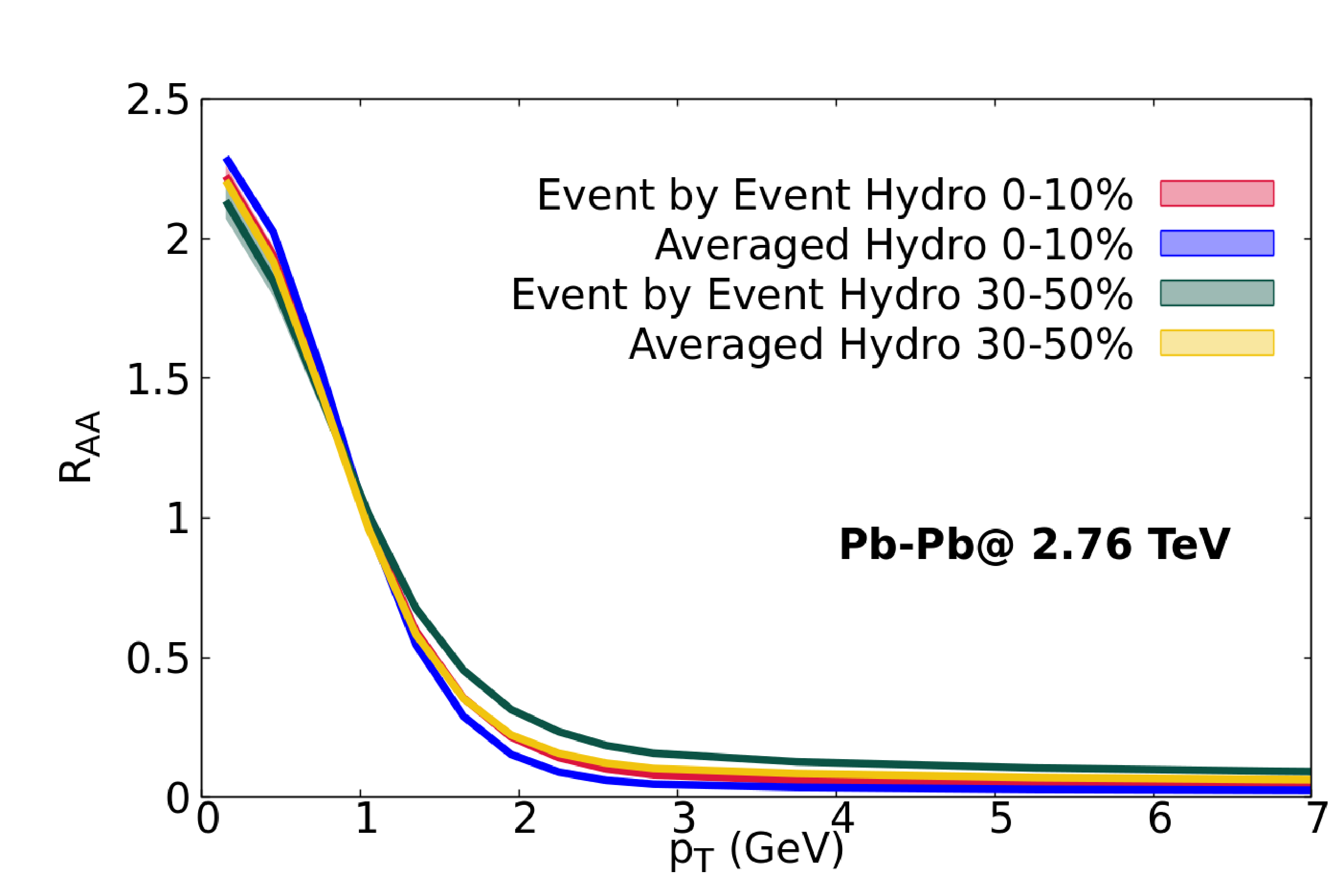}
    \caption{(Left panel) Updated charm quark diffusion coefficient (setup-II) as a function of temperature. (Right panel) $R_{AA}$ of D meson within the hybrid model for heavy quark using setup-II. }
    \label{fig2}
\end{figure*}
\subsection{Updated charm quark diffusion in hybrid model for charm quarks}
Heavy quark coupling in the QGP medium can be characterized by the spatial diffusion coefficient $D_s=\frac{T}{m_{HQ}A(p\rightarrow 0,T)}$, typically considered as a phenomenological parameter. We have observed that viscous effects have no visible impact on the temperature behavior of $D_s$. In this section, an updated charm quark diffusion coefficient has been considered, incorporating the following developments along with the elastic scattering process of heavy quarks in the medium:
\begin{itemize}
\item Soft gluon radiation: For the inelastic ($2\rightarrow 3$) process,
$HQ(P)+l(Q) \rightarrow HQ (P')+l(Q') + g(K')$,
with $K'\equiv(E_{k'},{\bf k'_{\perp}},k'_z)$ as the four-momentum of the emitted soft gluon by the heavy quark in the final state, 
the thermal averaged $F(|{\bf p}'|)$ takes the form as,
\begin{align}\label{Ap1}
    \langle \langle F(|{\bf p}'|) &\rangle \rangle =\dfrac{1}{\gamma_{HQ}}\frac{1}{2 E_p} \int \frac{d^3 {\bf q}}{(2 \pi)^3 2E_q} \int \frac{d^3 {\bf p'}}{(2 \pi)^3 2E_{p'}}  \int \frac{d^3 {\bf q'}}{(2 \pi)^3 2E_{q'}} \int \frac{d^3 {\bf k'}}{(2 \pi)^3 2E_{k'}} \ (2 \pi)^4 \nonumber\\&\times \delta^{(4)}(P+Q-P'-Q'-K')  \sum{|{\mathcal{M}}_{2\rightarrow 3}|^2} \nonumber\\&\times f_{g/q}(E_q) (1 \pm f_{g/q}(E_{q'})) \ (1 + f_g(E_{k'})) \theta_1(E_p-E_{k'}) \ \theta_2(\tau-\tau_F) \ F(|{\bf p}'|),
\end{align}
where $|{\mathcal{M}}_{2\rightarrow 3}|^2$ describes the matrix element squared for the radiative process~\cite{Abir:2011jb}. The theta function $\theta_1(E_p-E_{k'})$ imposes constraints on the heavy quark initial energy and $\theta_2(\tau-\tau_F)$ indicates that scattering time $\tau$ is larger than the gluon formation time $\tau_F$ (Landau-Pomeranchuk-Migdal Effect).

\item Fixing parameters: The infrared regulator and coupling constant that enters through the matrix elements are the sources of uncertainty in the estimation of the charm quark diffusion coefficient.  In this study, the Debye mass, which is the conventional choice of the infrared regulator, is substituted with a realistic hard thermal loop parameterization of the regulator. Additionally, the study employs an effective coupling constant that incorporates non-perturbative dynamics.
\end{itemize}
In Fig.~\ref{fig2} (left panel), we plotted the temperature behavior of $2\pi D_sT$.  For the pQCD elastic process, it is seen that $2\pi T D_s\approx 30-40$~\cite{Kurian:2020orp}, which is much higher that the recent $N_f=2+1$ lattice estimation~\cite{Altenkort:2023oms}, $N_f=0$ lattice data~\cite{Banerjee:2011ra} and quasiparticle model (QPM) result~\cite{Scardina:2017ipo}. It is observed that the inclusion of soft gluon radiation from heavy quarks suppresses $D_s$. This is because heavy quarks experience more drag force as they lose energy through collisional and radiative processes. The diffusion coefficient with radiative effects and updated parameters are described as setup-II in the figure. The revised infrared regulator and effective coupling seem to have prominent impacts on the temperature dependence of  $2\pi D_sT$.  We have developed a hybrid model for heavy quarks by incorporating the latest developments in the hydrodynamical description of QGP~\cite{Singh:2023smw}. In Fig.~\ref{fig2} (right panel), we plotted the nuclear suppression factor $R_{AA}$ of D-mesons utilizing the updated charm quark diffusion coefficient (setup-II) in the hybrid model. Results from event-by-event initialized calculations have been compared with those from smooth initial conditions. 
\section{Summary}
We have investigated the dynamics of heavy quarks and the associated transport coefficients in an expanding QGP medium by using hydrodynamical modeling.  A systematic analysis has been conducted to explore the viscous corrections to the momentum evolution of charm quarks in the evolving medium. We observe that the viscous effects exhibit weaker dependence on charm quark momentum evolution, particularly in the initial stages of the collision. We have studied the impact of heavy quark radiative process and non-perturbative effects on the diffusion coefficient. Further, we have estimated the nuclear modification factor of D-mesons with the updated diffusion coefficient by employing a hybrid model for charm quarks for Pb+Pb collision at 2.76 TeV.
\vspace{6pt} 

\funding{This research was funded by U.S. DOE Grant No. DE-FG02-87ER40328 and DE-SC-0024347, and Natural Sciences and Engineering Research Council of Canada under grant numbers SAPIN-2018-00024 and SAPIN-2020-00048.}

\acknowledgments{Numerical
computations were done on the resources provided by the
Minnesota Supercomputing Institute (MSI) at the University
of Minnesota and on Beluga supercomputer at McGill University managed by Calcul Québec and the Digital Research
Alliance of Canada. M .K. acknowledges the Special Postdoctoral Researchers Program of RIKEN. }

\conflictsofinterest{The authors declare no conflict of interest.}

\appendixtitles{no} 
\appendixstart
\begin{adjustwidth}{-\extralength}{0cm}

\reftitle{References}

\bibliography{hq_proceedings_mk}

\PublishersNote{}
\end{adjustwidth}
\end{document}